\input epsf.tex
\documentstyle[12pt]{article}

\newcommand{\be}{\begin{equation}}   \newcommand{\ee}{\end{equation}}
\newcommand{\bear}{\begin{eqnarray}}
\newcommand{\eear}{\end{eqnarray}}
\newcommand{\ba}{\begin{array}}      \newcommand{\ea}{\end{array}}

\newcommand{\ie}{{\it i.e.\ }}

\begin{document}
\pagestyle{empty}
\begin{titlepage}

\vspace*{-8mm}
\noindent 
\makebox[11.5cm][l]{BUHEP-97-18} hep-th/9710150 \\
\makebox[11.5cm][l]{UCTP 107/97} March 2, 1998 \\


\vspace{2.cm}
\begin{center}
  {\LARGE {\bf  The imaginary time Path Integral
and non-time-reversal-invariant- saddle points of the Euclidean Action  }}\\
\vspace{42pt}
Indranil Dasgupta \footnote{e-mail address:
dgupta@physics.uc.edu}

\vspace*{0.5cm}

 \ \ Department of Physics, University of Cincinnati \\
{400 Geology/Physics Building, Cincinnati, OH 45221, USA}

\vskip 3.4cm
\end{center}
\baselineskip=18pt

\begin{abstract}
{\normalsize { We discuss new bounce-like (but non-time-reversal-invariant-)
solutions to Euclidean
equations of motion, which we dub {\it {boomerons}}.
In the Euclidean path integral approach to quantum
theories, boomerons make an imaginary contribution to the vacuum
energy. The fake vacuum instabilty can be removed by cancelling boomeron
contributions against contributions from 
time reversed boomerons (anti-boomerons). The
cancellation rests on a sign choice whose significance is not completely
understood in the path integral method. }}

\end {abstract}

\vfill
\end{titlepage}

\baselineskip=18pt  
\pagestyle{plain}
\setcounter{page}{1}


\section {Introduction}

Formally the imaginary time path integral is related to the trace of the
imaginary time evolution operator ${\rm {exp}} (-Ht)$ by the relation:
\be
{\rm {lim}}\, _{t \to \infty}{\rm {Tr}} [{\rm {exp}}
 (-Ht/\hbar)] = N\int {\cal {D}} \Phi
{\rm {exp}} (-S_E[\Phi]/\hbar), 
\label {first}
\ee
where $\Phi$ denotes the coordinates in the
quantum theory, $H$ is the Hamiltonian operator, $S_E$ is the Euclidean
action (the imaginary time action) and $N$ is a
normalization factor. 
This formula is often exploited to compute
the decay rates of metastable false ground states of the quantum theory
by evaluating the path integral on the right semiclassically. Briefly,
the procedure \cite {coleman} can be understood by considering particle
mechanics in one dimension. The action in real time is $S=\int dt
[{1\over 2} \dot {q}^2 - V(q)]$. 
Consider a potential
$V(q)$ as shown in Fig. 1. The point $q=a$ is a classical ground
state. Quantum mechanically the particle penetrates the barrier to the
right. The tunneling probability can be computed by the WKB method. The
path integral (P.I.) method 
is an elegant alternative method, where one simply expands
the right hand side of (\ref {first}) about classical solutions of the
Euclidean equations of motion that extremize the imaginary time
action $S_E$ and satisfy the boundary
condition ${\rm {lim}}\, _{ t \to \pm \infty}\, q = a$. Since $S_E = 
\int dt [{1\over 2} \dot {q}^2 + V(q)]$, the extrema of $S_E$ are 
given by classical paths of a particle moving in the potential $-V(q)$. 

\bigskip
\centerline{\epsfxsize=3.0in\epsfbox{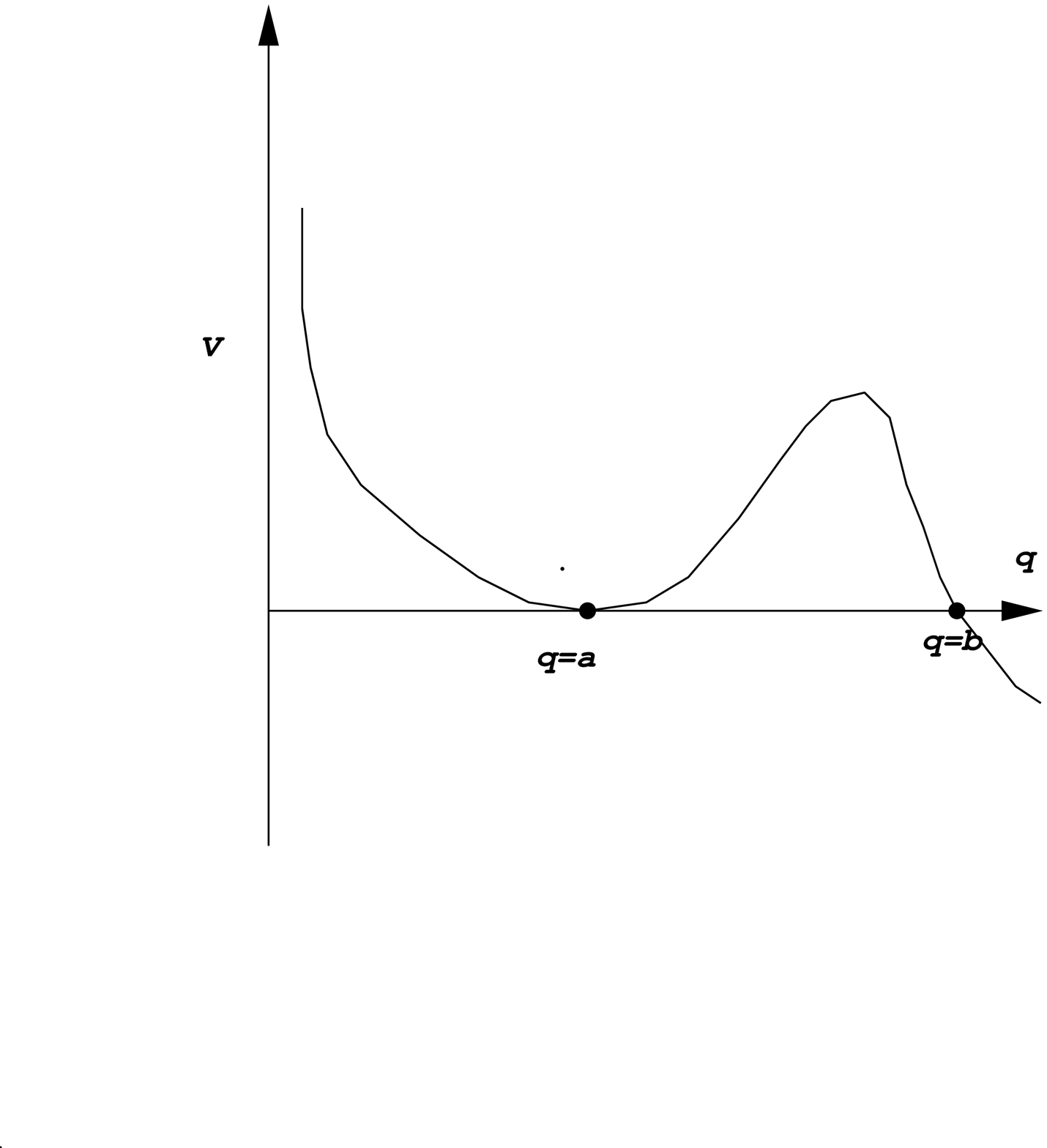}}
\vspace {-0.5cm}
\makebox[0.8in][l]{\hspace{2ex} Fig. 1.}
\parbox[t]{4.8in}{ {\small The potential $V$ has a local minimum at
$q=a$. The escape point is at $q=b$.}}

\bigskip

In this case there is a nontrivial classical
solution to the equations of motion known as the bounce. The bounce is
given by the path $\overline {q}(t)$ that begins at $q=a$ when $t \to
- \infty$, ``bounces'' off the point $q=b$ (with $V(b)=V(a)=0$) and ends at
$q=a$ as $t \to \infty$. Expanding $S_E$ about $\overline {q}$ up-to
second order in the fluctuations we get, $S_E[\delta q] = S_E[\overline
{q}] + \int dt\, \delta q [-d_t^2 + {d^2V \over dq^2}(\overline {q})]
\delta q + {\rm {higher\, order}}$, with $\delta q = q-\overline
{q}$. The functional integral
over the quadratic fluctuations is a simple Gaussian one if the
eigenvalues of the operator $O[\overline {q}] = -d_t^2 + {d^2V \over dq^2}(\overline
{q})$ are all non-negative. It is well known however, that the bounce is
a saddle point of the action $S_E$ and $O[\overline {q}]$ has a 
single negative eigenvalue. The ill defined 
integration over the negative eigenmode can be done by 
continuing the eigenmode $\psi$ to complex values, i.e., the integral
$\int d\psi \, {\rm {exp}}(-\alpha \psi ^2)$, where $\alpha$ is the negative
eigenvalue of $O[\overline {q}]$ can be performed through a contour along the
imaginary $\psi$ axis. The result is that the integral over the Gaussian
fluctuations yields a formally imaginary quantity: $[{\rm {det}}
(O[\overline {q}])]^{-1/2}$. When contributions from multi-bounce solutions of the
Euclidean equations of motion are taken into account, equation (\ref
{first}) reduces to \cite {coleman}:
\be
{{\rm lim\,}} _{t \to \infty}{\rm {Tr}} [{\rm {exp}} (-Ht/\hbar)] 
 = {\rm {lim\,}}_{t \to \infty} 
\left (\omega \over \pi \hbar \right )^{1/2} {\rm {e}}^{-\omega t/2}
{\rm {exp}}\left [ {\rm {exp}}(-S_E(\overline
{q})/\hbar) \times K t \right ][1 + {\rm {O}}(\hbar)],
\label {second}
\ee
where $K = (S_E[\overline {q}]/2 \pi \hbar)^{1/2} 
\left [{{\rm {det}}^{\prime}\, O[\overline {q}] \over {\rm
 {det}}\, O[a]} \right ]^{-1/2} $, $\omega = {d^2 V \over d
 q^2} {\Big |}_{q=a}$,
$O[a] = -d_t^2 +{d^2 V \over d q^2}(q=a)$ and the prime on one of
the determinants indicates that the zero eigenvalue of $O$ coming from
the time translational invariance of the bounce is to be excluded in the
computation of the determinant. 
Comparing this with (\ref {first}) one sees that the ground state energy
is given by $E_0 = {1 \over 2}\hbar \omega + \hbar K {\rm {exp}}(-S_E
[\overline {q}]/\hbar)[1 + {\rm {O}}(\hbar)]$. 
The first term in this expression is just the ground state energy of a
harmonic oscillator, but the second term 
is imaginary and may be interpreted as
the decay width of the metastable state localized near $q=a$. 
Although the imaginary term is formally
smaller in magnitude than order $\hbar ^2$ terms that are neglected in
this approximation, it is still the dominant contribution to any
imaginary part in the energy and, heuristically, should provide an 
estimate of the decay width.

The confidence in the above method comes from its ability to
reproduce the results of the more rigorous WKB approximation where 
the tunneling is seen as happenning through the path of
least resistance $\overline {q}(t), t \in (-\infty, 0)$
that extends between $q=a$ and the ``escape point''
$q=b$ on the ``other side'' of the
barrier. 
The bounce solution of the
P.I. approach is exactly the ``sum'' of the least resistance path
of WKB and its time reversal conjugate. { \it {The bounce itself
therefore, is time-reversal-}} (${\cal {T}}$) 
{ \it {invariant}}.

This formalism can be extended to multiple
dimensions \cite {wkb}. In a system
with $n$ coordinates $q_i$, where $i=1,..n$, the paths of
least resistance connect the false vacuum $q_i^f$ to a surface $\sigma$
of zero potential that lies on the ``other side'' of the barrier. 
In the P.I. formalism, each such path extends to a
${\cal {T}}$ invariant bounce and makes an imaginary contribution to
the vacuum energy. If there are several saddle points $\overline
{q}_{i}(\alpha)$ ($\alpha = 1,2..m$), the imaginary
part of the vacuum energy is given by 
\be
\delta E_0 = \hbar \sigma _{\alpha =1}^m K_{\alpha} {\rm
{exp}}(-S_E[\overline {q}_i(\alpha)/\hbar])[1 + {\rm {O}}(\hbar)] \, .
\label {multisaddle}
\ee
In this approximation one need retain only the 
dominant bounces which are the bounces with the least (Euclidean) action.

While the saddle points in the P.I. formalism include all the WKB
trajectories as bounces, it is not obvious 
whether the least action saddle point of $S_E$ 
is always a bounce.
If in some system the dominant imaginary contribution to the
ground state energy comes from saddle points that are not bounces,
the corresponding ``decay width'' (if non-zero) has nothing to do with
tunneling and will signal a limitation of the P.I.
formalism. 
In a non-compact 
one dimensional system like the one shown in Fig. 1, all periodic
solutions of the Euclidean equations of motion must retrace themselves
backwards and be ${\cal {T}}$ invariant.
This need not be true in systems with more degrees of freedom.
Consider a two
dimensional system with coordinates $q_1$ and $q_2$ (Fig. 2). The origin is a
minimum of the potential and there may be a loop-like 
trajectory along $P$ that extremizes $S_E$ but is not invariant under
time reversal. 

\centerline{\epsfxsize=3.0in\epsfbox{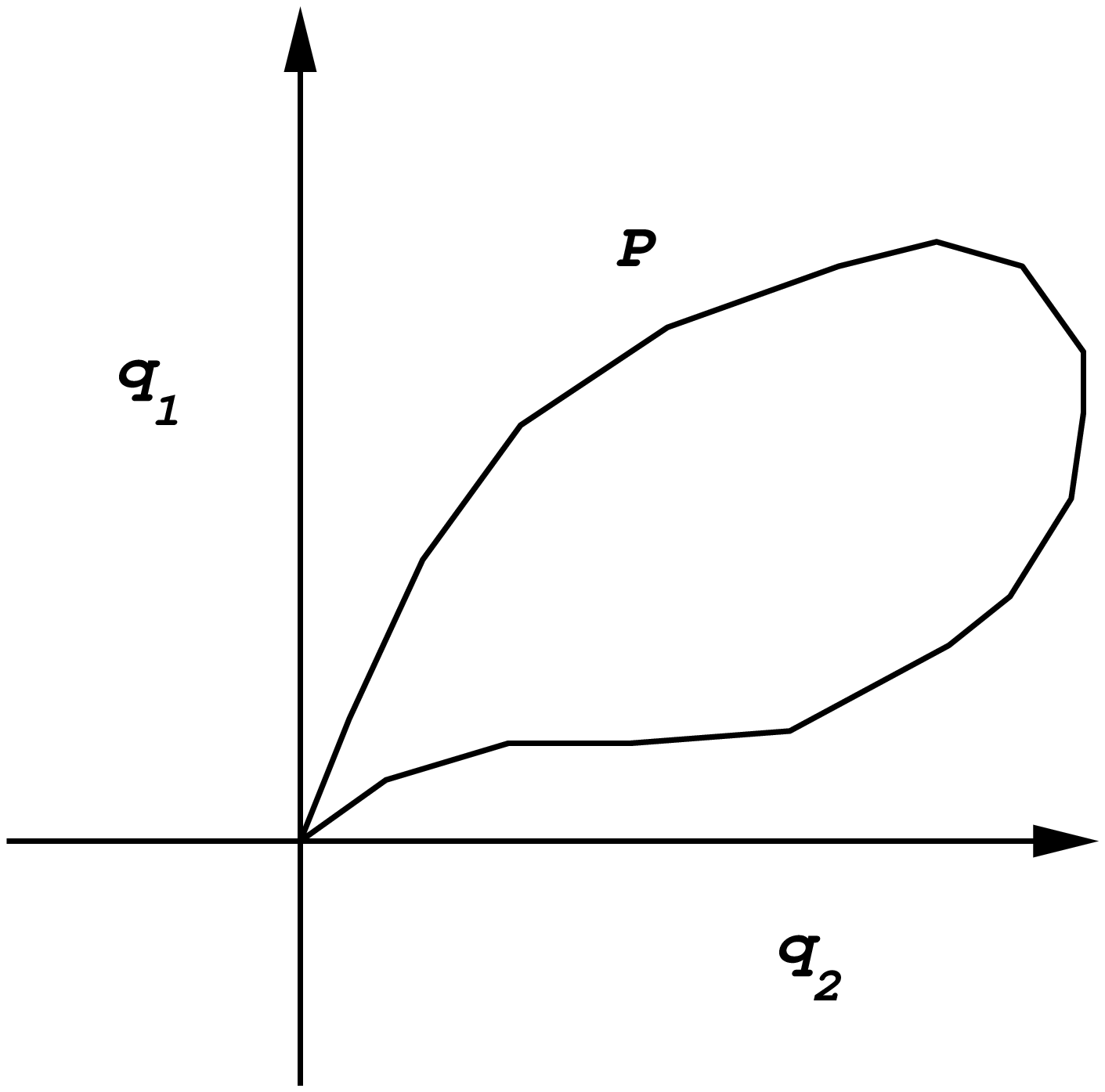}}

\makebox[0.8in][l]{\hspace{2ex} Fig. 2.}
\parbox[t]{4.8in}{ {\small There is a trajectory along the
 path $P$ that is not invariant under time-reversal but
extremizes $S_E$. 
} }
\bigskip

Because they lack ${\cal {T}}$ invariance, trajectories like this
are not bounces and there is no corresponding WKB trajectory.
Nevertheless, in the P.I. formalism, the
determinant factor in $K$ computed for them can in principle have a single
negative eigenvalue and, like a bounce, they can make an imaginary contribution to the
vacuum energy.

At this point the following questions arise:\\
(i) Do trajectories like this actually exist for which $K$ is imaginary? \\
(ii) Can such fake bounce solutions exist in quantum field theories?\\
(iii) Does the P.I. formalism for computing the decay width of
the vacuum fail when such solutions exist?

We will find answers to these questions in the next sections. In
particular we will find that 
such fake bounces can exist even in field theories and indicate how
one can take their effect into account in the P.I.
formalism. Because these solutions to the equations of motion resemble
(allegedly possible) periodic trajectories of a boomerang, we will
call them ``boomerons'' to facilitate distinction with the bounces.

\section {Boomerons in Quantum Mechanics }

Let us show that the answer to question (i) of the previous section is
in the affirmative. 
Consider quantum mechanics for a particle of unit mass moving on a $2$ dimensional
sphere ($S^2$). Suppose the potential energy is minimized at the north pole
where the particle attains its classical ground state.
We are interested in closed trajectories on the
sphere with one point fixed at the north pole. The space ${\Omega (S^2)}$ of
these
trajectories is not simply connected. In the language of algebraic geometry, 
$\Pi _1[\Omega (S^2)] \simeq Z$, where $Z$ is the set of all integers (see, for
instance ref.~\cite {algebra}).

If $\Omega (S^2)$ were a compact and finite dimensional space, this fact 
alone would suffice to prove the exitence of a boomeron. This is best 
illustrated by Fig.~3a where the torus is a compact and finite dimensional 
analogue of $\Omega (S^2)$. The height function on the torus plays the role
of $S_E$. If we look at non-contractible curves of winding number 1 based at 
the global minimum of $S_E$ at $P$, the saddle point at $B$ emerges as the 
highest point of the curve with the lowest highest point (the mountain pass 
curve). The argument breaks down even for a simple non-compact space like the
tapered cylinder of Fig.~3b where the saddle point $B$ ``escapes'' to 
infinity. In such cases, neither the existence nor the ${\cal {T}}$- non- invariance
of the saddle point can be guaranteed. Nonetheless, 
the non-triviality of $\Pi _1 [\Omega (S^2)]$ is an encouraging signal for
the boomeron hunter. In this particular 
example the boomeron can be explicitly constructed for simple potentials on 
$S^2$, such as a potential $U(\theta)$ that is invariant for rotations around the axis
joining the north and south poles ($\theta $ is the latitude). 
The boomeron trajectory $B$ then traces out a large circle passing through the north pole (Fig.~4).
It is simple to show, when $U(\theta)$ increases monotonically toward the south pole,
that the Hessian operator $O$
computed at the boomeron trajectory has a negative eigenvalue (roughly,
this unstable direction corresponds to sliding off the trajectory to the left or right,
so that it moves into the left or the right hemisphere defined by the boomeron).
The action of the boomeron is $S_E = \int d \theta \sqrt {{R^2U(\theta) \over 2}}$, where 
$R$ is the radius of the sphere. The corresponding
boomeron contributions to the ground state energy are imaginary and of order $({S_E \over 2\pi \hbar})^{1/2}
{\rm {exp }}(-S_E/\hbar)$. 

\vspace {1cm}
\centerline{
\epsfxsize=2.0in \epsfbox{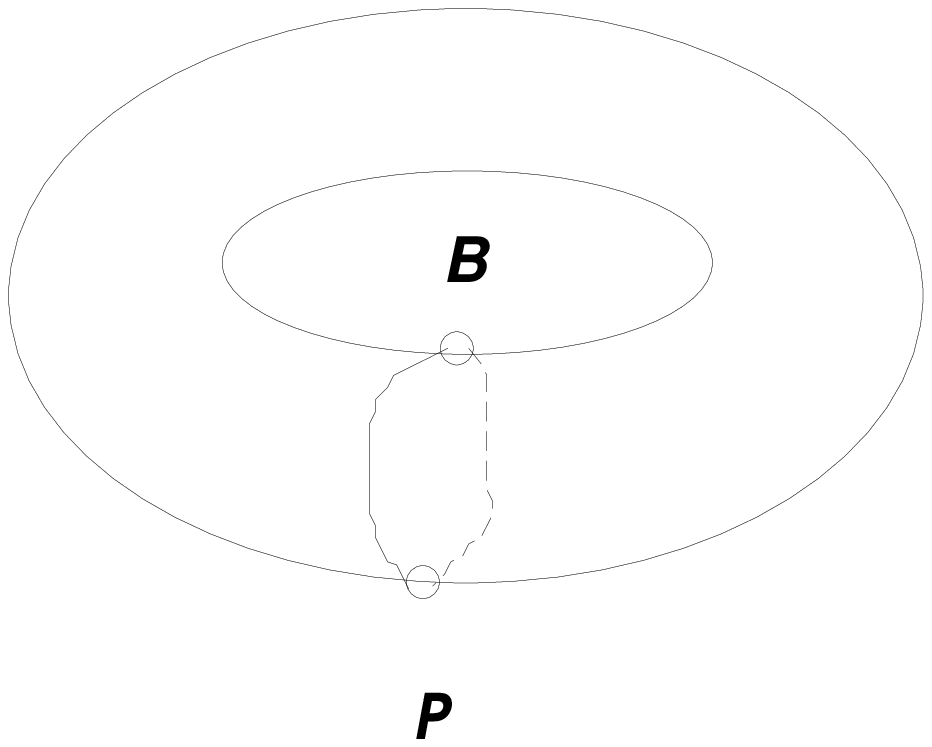} 
}

\noindent
\makebox[0.8in][l]{\hspace{2ex} Fig. 3a.}
\parbox[t]{4.8in}{ {\small $P$ is the minimum of the height function on the 
torus. Of a set of non-contractible loops through $P$, the
saddle point $B$ is the highest point on the loop with the lowest highest
point.} }
\bigskip

\centerline{
\epsfxsize=2.0in \epsfbox{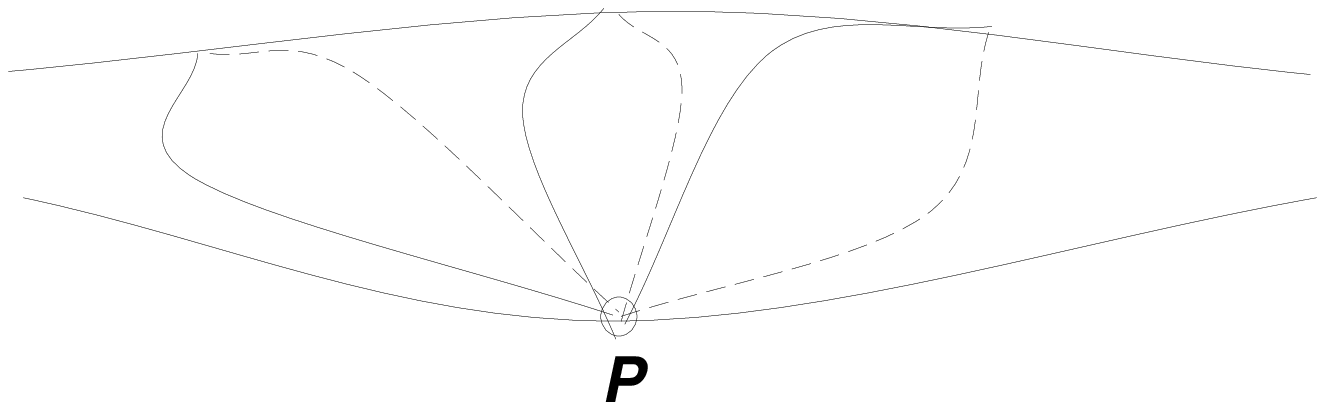}
}

\noindent
\makebox[0.8in][l]{\hspace{2ex} Fig. 3b.}
\parbox[t]{4.8in}{ {\small Three non-contractible curves are drawn on an 
open cylinder. The noncontractible curve
through $P$ with the lowest highest point ``escapes to $\infty$'' toward 
left and right.
} }

\bigskip

\centerline{
\epsfxsize=2.0in \epsfbox{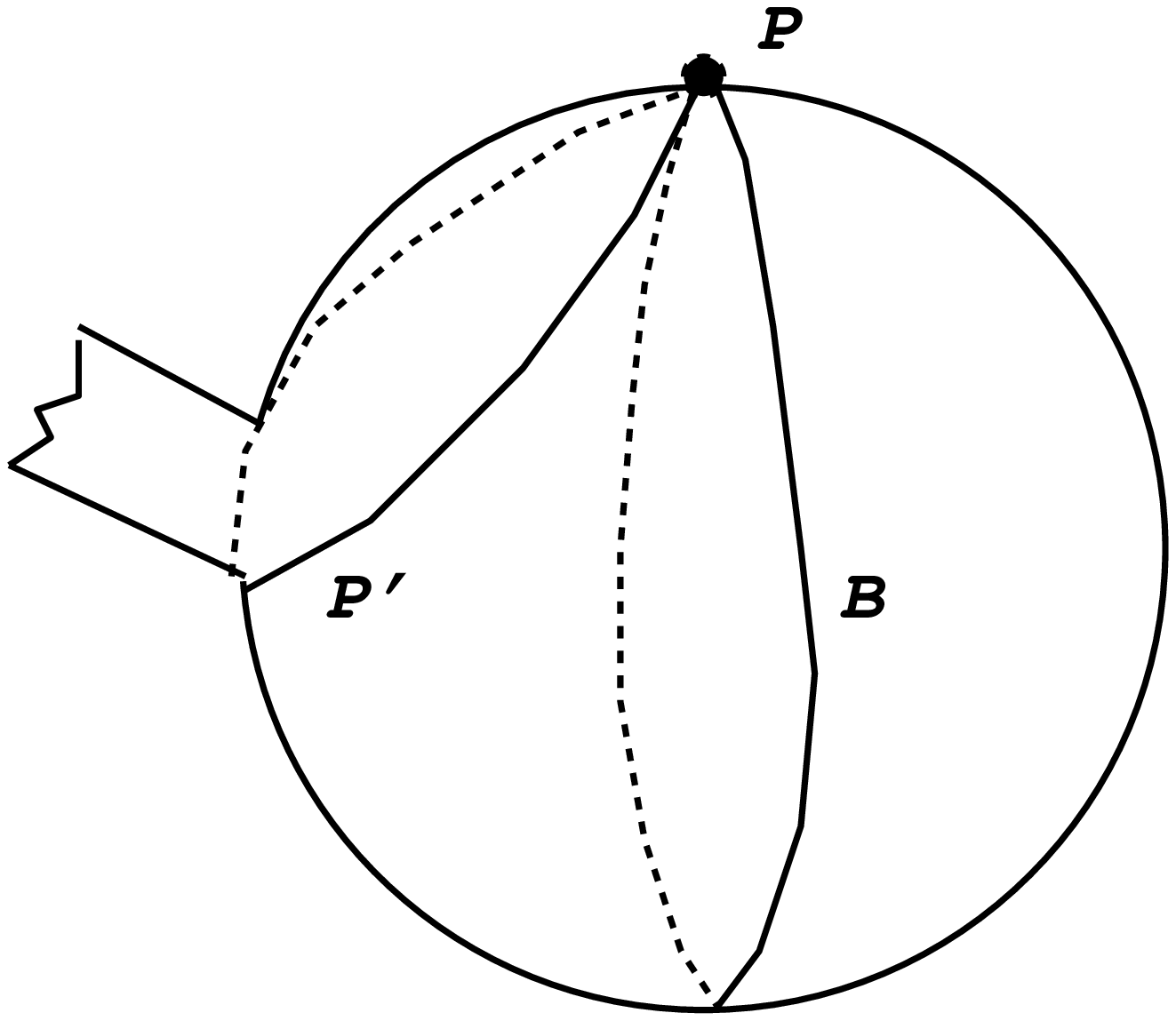}
}

\noindent
\makebox[0.8in][l]{\hspace{2ex} Fig. 4.}
\parbox[t]{4.8in}{ {\small Attaching an infinite cylinder to the sphere
creates a new minimum $P^{\prime}$ of $S_E$, while the boomeron $B$ may
continue to exist. } }

\bigskip

This example not only establishes
the existence of boomerons in quantum mechanics but 
also sheds light on an interesting aspect of the boomerons:
they are likely to exist whenever the classical vaccum is the
global minimum of the potential and 
the space of closed trajectories passing
through the classical vacuum 
is not a simply connected space. 
We do not preclude the existence boomerons that are not ``topologically
required'' as above. For instance by attaching an infinite cylinder to the sphere
(Fig. 4), the topology of the configuration space is changed from $S^2$ to 
$R^2$. Since $\Pi _1[\Omega (R^2)]$ is trivial, there are no topological 
requirements for the existence of the boomeron. Nevertheless, the boomeron
continues to exist if the attached
cylinder does not intersect the boomeron's path. In this case however, a
new minimum of the action must appear in the form of the trajectory
$P^{\prime}$. The boomeron $B$ is the saddle point between the two
action minima at $P$ and $P^{\prime}$.

\section {Boomerons and Path Integrals}

Before we move on to examples of boomerons in field theories
let us attempt to answer the question posed in (iii) of section 1
at the level of quantum mechanics.
Note that a boomeron is a nuisance in the P.I. formalism only
if its action is lower than the action of all
bounces. It is not hard to construct theories where this is
true. Indeed, in the example constructed in section 2, bounces do not
not exist at all while boomerons do. 
In these theories, at first glance, the P.I. formalism seems to fail.
However, there is a simple way of saving the
method. Not surprisingly it rests on the 
single fact
distinguishing boomerons from bounces, namely, the boomeron's non-invariance
under ${\cal {T}}$. The time reversed boomeron is a different
boomeron. One could give it the suggestive name anti-boomeron and ask if
its contribution can cancel the contribution of the
boomeron. 

This does appear to be a possibility in the P.I.
method. Recall that the imaginary term $K$ in (\ref {second}) suffers
from a sign ambiguity. When
carrying out a formally divergent Gaussian integral of the kind $\int d\psi
{\rm {exp }}(-\alpha \psi ^2)$ with a negative $\alpha$ the analytic
continuation of $\psi$ to
imaginary values can be done in two ways. Integrating $\psi
$ from $-i\infty$ to $i\infty$ and from $i\infty$ to $-i\infty$ 
give opposite signs for
the integral. In a theory with no boomerons,
the sign choice for the imaginary part of the vacuum energy
does not change the physical value of the decay width. When both
boomerons ($B$) and anti-boomerons ($\overline {B}$) 
are present and dominant, their imaginary 
contributions to the vacuum energy can be exactly cancelled against each
other if the
action is invariant under time reversal ($S_E[B] =
S_E[\overline {B}]$)\footnote { We discuss theories that are
not invariant under time-reversal in a later section.}. One does this
simply by choosing 
opposite signs for $K$ in the boomeron and the anti-boomeron in equation
(\ref {multisaddle}). 

What is unsatisfactory about this otherwise simple and straightforward 
procedure is that the
P.I. formalism does not {\it {require}} us to make this
sign choice. This is not a disaster, since 
the P.I. method for computing vacuum decay widths 
is, despite its elegance, a heuristic method.
In the presence case it is not so much the wrong result that one
gets upon a ``wrong'' choice of signs, as the 
possibility of a perfect cancellation
with the ``correct'' sign choice that we find intriguing. One would
hope to understand the reason for this at a deeper level, perhaps in an
extended version of the P.I. method where the
orientation of the analytic continuation at one saddle point
fixes the orientation of the analytic continuation at other saddle
points in an un-ambiguous manner. 

There is also the need to check the familiar
regularizations of the P.I. method carefully to see if the boomeron and
the anti-boomeron contributions {\it {can}} indeed have opposite signs. 
This is actually a tricky point which we would like to clarify
with an example.
Suppose a boomeron exists in a 4-
dimensional bosonic field theory with real fields $\phi_1,
...\phi_n$ which are components of scalar or gauge fields. 
The boomeron $\phi_i (x,y,z,t)$, 
is not invariant under time reversal ($\phi_i (x,y,z,t) \ne \phi_i
(x,y,z,-t)$). The finiteness of the boomeron action demands that the
fields should approach constant values as $x,y,z,t \to
\infty$. Applying an $SO(4)$ transformation $\Lambda$ to the above configuration
preserves the boundary conditions and 
yields another field configuration $\Lambda \phi_i$ which is also a
boomeron with the same action. 
Integrating over the zero mode of $SO(4)$ transformations therefore amounts
to adding these boomeron contributions with the {\it {same sign}}. This is
{\it {required}} in any regularization since 
the boomerons are connected by a 
continuous zero mode but the {\it {sign}} can not change continuously. 

On the other hand, 
the anti-boomeron $\overline {\phi_i}$ and its $SO(4)$ transforms ($\Lambda
\overline {\phi_i }$) are obtained from the original boomeron $\phi_i$ 
by applying an $O(4)$ transformation 
(namely, ${\cal {T}}$) that
is {\it {not connected to identity}}. 
Thus the boomerons and anti-boomerons
remain separated in the space of field configurations, and there is no
apparent obstruction to having opposite analytic continuation
at boomerons and anti-boomerons. The fact that 
$O(4)$ is not a connected group is vital. The same is
true of the symmetry group $O(d)$ in $d$ dimensional field theories ($d
\ge 2$). When $d=1$, the appropriate symmetry group is $Z_2$
(${\cal {T}}$) which is also a
discrete group.

Although there is no obstruction to choosing opposite signs from continuous space-time
symmetries relating boomerons and anti-boomerons, there may be continuous internal symmetries
relating them. Once again an example will explain this situation best. In the 2- dimesional
quantum mechanics described in the previous section, the boomeron and the antiboomeron 
trajectories are coincident large circles with opposite orientations. Because the potential
$U(\theta)$ has a rotational symmetry in the $\phi$ (the longitude) direction, one can obtain
the anti-boomeron from the boomeron by continuously changing $\phi$ to $\phi + 
\pi $ through a zero mode of boomerons.

The complication due to internal symmetries will also arise in field theories, especially
in the form of gauge-symmetries. Here we suggest a regularization that may be used to preserve
the boomeron- anti-boomeron cancellation. One simply adds a term $\delta U$ to the potential
that breaks the discomforting continuous symmetry but preserves time-reversal-invariance
of the action. Then the boomeron- anti-boomeron imaginary parts can cancel. The $\delta U \to 0$
limit is taken after the cancellation. Physically, if the cancellation gives the correct result
in the regulated theory, it should be valid in the symmetric theory. 

In higher than 2- dimensional- 
gauge theories, regularizations
that break gauge symmetries may spoil the renormalizability of the theory. However, the zero mode
relating the boomeron and the anti-boomeron is an infra-red effect and its resolution should 
be independent of the high energy behaviour of the theory. In this spirit, one can define
a cut-off theory by integrating out all the high energy modes. In the low energy effective theory,
the term $\delta U$ may be added (it can be thought of as coming from a renormalizable Higgs
interaction, with a massive Higgs field that has been integrated out). The subsequent 
regularization of boomeron contributions may proceed as sketched above. Having made our suggestion
we remark that we will not demonstrate this procedure by working out an example in detail.
This is not for lack of examples in gauge theories, but as we shall see shortly, all
the gauge boomerons we could find, offer other (peculiar but easier) solutions to the symmetry
problems described here.
In the next section we consider in greater detail these peculiarities of boomerons
in field theories.

\section {Boomerons in Field Theories}

\subsection {Scalar Fields}

It is simple to show that for purely scalar field theories in dimensions
greater than 2, the existence of a boomeron implies 
the existence of a bounce with an equal or 
lower action.
Consider a purely scalar field theory in $d$ ($d \ge 2$) 
dimensions. The Euclidean action is
\bear
S_E &=& T + V \; , \nonumber \\
T &=& \int d^dx {1 \over 2}\left [ \left ( {d \phi_i \over d x_{\mu} } \right )^2
\right ] \; , \nonumber \\
V &=& \int d^dx \left [ U(\phi_1, ...\phi_n) \right ] \; ,
\label {scalaraction}
\eear
where $\phi_i$ are $n$ real scalar fields and $U$ is a potential { \it {which
is normalized to be zero at the vacuum of interest}}. Under
scale transformations $x \to \lambda x$ with the positive parameter
$\lambda $ the terms $T$ and $V$ scale like: $T \to \lambda ^{d-2} T \;
; V \to \lambda ^d V$. At a stationary point $\overline {\phi }_i(x)$ 
of the action we must therefore have,
\be
{d S_E [\overline {\phi}_i] \over d \lambda }{\Big |}_{\lambda =1} = (d-2)
T[\overline {\phi}_i] + (d) V[\overline {\phi}_i] = 0 \; .
\label {lambda}
\ee
This condition is satisfied by a non-trivial finite action field configuration only
if $V \le 0$, which implies that the vacuum of interest is a false vacuum
(\ie $U$ becomes negative somewhere) and
bounce solutions also exist. 

It is straightforward to show that there is always a bounce with an
equal or lower action than any boomeron (should a boomeron exist). 
Suppose a boomeron $\overline {\phi_i}$ 
exists with $T$ and $V$ obeying (\ref {lambda}). We
can slice the boomeron into two halves using a $d-1$ dimensional
plane. Let us call the two halves $L$ and $R$ (for left and right
halves respectively). 
The integrals $T$ and $V$ appropriately split into integrals over
$L$ and $R$ given by $T(L), T(R), V(L)$ and $V(R)$ respectively with
$T=T(L) + T(R)$ and $V=V(L) + V(R)$. We slice the boomeron so that $T(L)
= T(R) = {1 \over 2}T$. The different ways of slicing the boomeron
while preserving this condition are in one to one correspondence with the
points on the group space $SO(d)$. We choose a slicing for which $V(R)$ is
least. Then $V(R) \le V(L)$. Let us call the axis perpendicular to
the $d-1$ dimensional plane separating $L$ and $R$ the time axis
$t$. The slicing plane
intersects this axis at $t=0$ by convention, with $t>0$ in the right half.

Using equation (\ref {lambda}), 
the action of the boomeron is $T[\overline {\phi} _i]+V[\overline {\phi} _i] = {2 \over d} T[\overline {\phi} _i]$. A
time- reversal- invariant- field
configuration $\hat {\phi} _i$ 
is obtained from the boomeron $\overline {\phi}_i$ by replacing the 
left half with a mirror image of the right half. 
\be
\hat {\phi} _i(x,...t) = \left\{ \begin {array} {rcl}  \overline {\phi}_i(x,... -t)
\; \; &{\rm {for}}& \; t \le 0 \\ [3mm] 
\overline {\phi}_i(x,...\, t) \; \; \; \; \; \; &{\rm {for}}& \; t > 0 \; \end {array}
\right.
\label {phihat}
\ee
The action of the new field configuration is lower than or equal to the
the action of the boomeron
\be
S_E[\hat {\phi}_i] = 2 \left [ T(R)[\overline {\phi}_i] +
V(R)[\overline {\phi}_i] \right ] \le {2 \over d} T[\overline {\phi}_i]
\; .
\label {phihataction}
\ee
Although $\hat {\phi}_i$ is not likely to be smooth at $t=0$, there are
smooth field configurations in the neighbourhood of $\hat {\phi}_i$
whose action is as close to that of $\hat {\phi}_i$ as one wants. The
configuration $\hat {\phi}_i$ does not satisfy (\ref {lambda}) but one
can scale transform the configuration by the parameter $\beta $ ($
x, ..t \to \beta x, ... \beta t$) so that condition (\ref
{lambda}) is satisfied. Then the following statements are true for the
scale transformed configuration $\hat {\phi }^{\beta }_i$.

(i) $\beta \le 1$.\\
 We show this as follows. Let us write $T =
T[\overline {\phi}_i]= T[\hat {\phi}_i]$ and $V = V[\hat
{\phi}_i]$. 
We scale transform $T$ and $V$ so
that $ (d-2)(\beta ^{d-2} T) + (d) (\beta ^d V) =(d-2)(T[\hat
{\phi_i^{\beta}}]) + (d)(V[{\phi_i^{\beta}}]) =0$. 
This condition, along with the condition, $V \le  {2-d\over d} T$, 
implies that $\beta
\le 1$. 

(ii) $S_E[\overline {\phi}_i] = {2 \over d} T \ge S_E[\hat
{\phi}^{\beta}_i] = \beta ^{d-2} T + \beta ^d V$.\\
Define $\alpha$ by the relation $V = - \alpha T$. 
Then, ${2 \over d}T \ge \beta ^{d-2} T + \beta ^d V = \beta ^{d-2} 
[1-\alpha \beta ^2 ]T$, iff, $ \beta ^{d-2} \le 1$ ($\alpha \ge {d-2 \over
d}$). Therefore the
assertion in (ii) follows from the assertion in (i). 

$\hat {\phi}^{\beta}_i$ is therefore a field configuration which
satisfies (\ref {lambda}) and is invariant under time- reversal. It has
been shown elsewhere \cite {dg, wipf} 
that the action of such a field configuration bounds the action of a
bounce from above. By the result in (ii) 
a bounce exists whose action is equal to or lower than the action of the
boomeron. 

The above results are actually valid for $d=2$ also. However one must
remember that the action in equation~(\ref {scalaraction}) does not include
the most general renormalizable terms when $d=2$. In particular,
non-linear sigma models are excluded from the discussion. We will
come back to this point again in the next section.

\subsection {Gauge and Topological Boomerons}

When the field theory has gauge fields in addition to scalar fields, the
arguments presented in the preceeding section do not hold.
Let us show that
in this case a boomeron may exist even if no 
bounce exists in the theory. We begin with a 
simple gauge theory in $d$ dimensions. The action is
\bear
S_E &=&T + V + F \; , \nonumber \\ 
T &=& \int d^dx \left [ |D_{\mu} \phi_i|^2 \right ]\; , \nonumber \\
V &=& \int d^dx \left [ U(\phi_1, ...\phi_n) \right ] \; , \nonumber \\
F &=& \int d^dx {1 \over 4} \left [ F_{\mu \nu}^2 \right ] 
\label {gaugeaction}
\eear
where $D_{\mu}$ is the covariant derivative and $F_{\mu \nu}$ is
the field strength. The potential $U$ is gauge-invariant.
One defines scale transformation by $\beta $ as
$(x,...t) \to \beta (x,...t), \; A_{\mu} \to {1 \over \beta}
A_{\mu}$. Under this transformation $T, V$ and $F$ transform as $T \to
\beta ^{d-2} T, \; V \to \beta ^d V, \; F \to \beta ^{d-4} F$. At a
stationary point of the action, we have an equation analogous to (\ref
{lambda})
\be
(d-2) T + (d) V + (d-4) F =0.
\label {beta}
\ee
This equation may have non-trivial solutions for $V \ge 0$ 
provided $d \le 4$. The dimension 
$d=4$ is a critical dimension where equation~(\ref {beta}) may
have non-trivial and finite action solutions only for $V=T=0$ (a
possibility, for instance, in pure Yang-Mills theories). 
Thus stationarity with respect to scale transformation 
produces no obstruction to boomerons 
in {\it {2, 3 or 4 dimensional gauge theories}} even if there is no vacuum
tunneling in the system. 

Indeed, boomerons in gauge theories may exist in the 
literature! The well known {\it {sphaleron}} \cite {km} of a
4- dimensional $SU(2)$ gauge theory is likely to be a boomeron in a 3- dimensional
$SU(2)$ gauge theory. Recall that the 4- dimensional theory has the action
\be
S_E = \int d^4 x \left [ {1 \over 4} F_{\mu \nu}^aF_{\mu \nu}^a
 + |D_{\mu}\Phi |^2 +
\lambda (\Phi ^{\dagger} \Phi - {1 \over 2} v^2)^2 \right ]
\label {sphalaction}
\ee
where the Higgs field $\Phi$ is a doublet under $SU(2)$, $F_{\mu
\nu}^a = \partial _{\mu} W_{\nu}^a - \partial _{\nu} W_{\mu}^a 
+ g \epsilon ^{abc} W_{\mu}^b W_{\mu}^c$ 
is the $SU(2)$ field strength, $W_{\mu} = W^a_{\mu}\tau _a$ are the $SU(2)$
gauge fields defined using the Pauli matrices $\tau _a$ and the covariant derivative
is $D_{\mu} = \partial _{\mu} - {1 \over 2} igW_{\mu}$. The sphaleron solution is an 
$SO(3)$ invariant non-trivial time- independent solution of the equations of motion
described by two real functions $f(\xi)$ and $h(\xi)$ (we follow the conventions
of ref.~\cite {km}) of the dimensionless coordinate $\xi = vgr$ where $r^2 =x_1^2 + x_2^2 + x_3^2$.
The gauge and Higgs fields are given by:
\bear 
W_{\mu} &=& - {2i \over g} f(\xi) (\partial _{\mu} U^{\infty})(U^{\infty})^{-1} \nonumber \\
\Phi &=& {v \over \sqrt {2}} h(\xi ) U^{\infty} \left (\begin {array} {c}  0 \\ 1 \end {array}
\right ) \, 
\label {sphalform}
\eear
where the $SU(2)$ elements $U^{\infty}$ are defined as $U^{\infty} = 
{i{\vec {\tau}} \cdot \hat {\bf {x}}}$. 

The sphaleron is a saddle point of energy and has at
least one unstable direction (a negative eigenvalue) associated with
it. In ref.~\cite {sphaleron} it is shown that the sphaleron has in fact
a single unstable direction when spherically symmetric perturbations are 
considered, provided $M_H <
12.03~M_W$, where $M_H$ is the Higgs mass and $M_W$ is the mass of the gauge
field. One expects that non-spherically symmetric perturbations will
increase the energy of these configurations. If this is true, then the
sphaleron in these cases is the lowest energy saddle point 
in the theory and, from our point of view, a boomeron in a 3- dimensional
theory where the Euclidean action is the 
same as the energy functional of the 4- dimensional theory. However
there is no false vacuum in the 3- dimensional theory and the question of bounces does
not arise. 

The sphaleron is a topological boomeron in essentially the same way as
our example from quantum mechanics in section 2 \cite {sphaltop, km}. One often 
says that
the existence of the sphaleron is related to the presence of
multiple topological vacua in the 4- dimensional theory. The vacua are
actually physically indistinguishable, being equivalent upto a gauge
transformation. However there are 
non-contractible trajectories in the space of field configurations
connecting these vacua. Because the end points are gauge equivalent,
they are very much like the loops in the space $\Omega (S^2)$ described in 
section 2.

The topological vacua correspond to
distinct maps from the compactified three space
$S^3$ to the gauge group $SU(2)$.
Such maps are characterized by the
integer winding number $n$. The 
sphaleron is a field configuration of winding number
$1/2$ which is the ``lowest'' saddle point of energy between the
vacua of winding numbers
$0$ and $1$ \cite {km}. 
In the 3- dimensional theory, where the sphaleron is a boomeron, 
a semiclassical estimate of the imaginary
part of the vacuum energy will be zero only if the contributions of the
sphaleron and the anti-sphaleron (whose winding number is $-1/2$) are
made to cancel by design. This issue is complicated by the fact that 
the sphaleron and the anti-sphaleron are actually gauge equivalent \footnote {
I am grateful to F. Klinkhamer for showing me this equivalence.}. It is instructive to
see how the gauge equivalence is established. In the 3- dimensional theory, we 
identify the coordinate $x_3$ as Euclidean time. Then
the sphaleron fields ($W, \Phi$) and the
anti-sphaleron fields ($\bar {W}, \bar {\Phi}$) are related by:
\bear
\bar {W}_i (x_1,x_2,x_3) & = & \left \{ \begin {array} {rcl} 
W_i (x_1, x_2, -x_3) \; \; & {\rm {for}} & \; i=1,2 \\ [3mm]
- W_i (x_1, x_2, -x_3) \; \; & {\rm {for}} & \; i=3 \end {array} \right. \nonumber \\
\bar {\Phi} (x_1, x_2, x_3) & = & \Phi (x_1, x_2, -x_3)  \; .
\eear
Therefore the anti-sphaleron fields are given by similar expressions as (\ref {sphalform})
with $U^{\infty}$ replaced by $\bar {U}^{\infty}(x_1, x_2, x_3) = {U}^{\infty}(x_1, x_2, -x_3)$. 
However $-\bar {U}^{\infty} = -i \tau ^3  {U}^{\infty} i \tau ^3$. Therefore, a global $SU(2)$
transformation by $-i \tau ^3$ takes ${W}$ to $\bar {W}$, while $\bar {\Phi}$ is obtained
from $\Phi$ by the same transformation along with a $U(1)$ charge rotation by ${\rm {exp}}
(i{\pi \over 2})$ (\ie an extra $i$). 

A peculiarity of the sphaleron as a boomeron can now be seen as follows. As we have
said before, a saddle point of the Euclidean action can be interpreted as a boomeron
only if it has trivial boundary conditions (all fields approach constant values at infinity). The 
sphaleron ansatz of (\ref {sphalform}) does not have this property. But there are 
no topological restrictions to choosing a different gauge in which $\Phi (x \to \infty) = 
{v \over \sqrt {2}}\left (\begin {array} {c}  0 \\ 1 \end {array} \right ) \,$ ; and
$W (x \to \infty) = {-2i\over g} (dU)(U)^{-1}$ where $U$ is the identity matrix ($U\equiv 1$).
In such a gauge
it is possible to define a fractional charge of the sphaleron by integrating the Chern-Simons
action density: $Q = {g^2\over 32 \pi ^2}\int d^3 x [F \wedge W - {2 \over 3}
W \wedge W \wedge W]$ \cite {km}. The value of $Q$
for the sphaleron in this gauge is $1/2$. One is now interested 
in continuous symmetry transformations that preserve the boundary conditions. Global symmetry
transformations that act non-trivially on $\Phi$ clearly do not preserve the boundary conditions.
This leaves us with the small gauge transformations (those, that
can be continuously obtained from the identity and do not change the boundary conditions).
However, small gauge transformations do not 
change $Q$, yet the charge $Q$ for the anti-sphaleron must be $-1/2$ as is easily seen
by applying time reversal to the form of $Q$. Because the sphaleron and the anti-sphaleron
are gauge equivalent the gauge transformation relating them is a large gauge transformation 
which can not be continuously obtained from identity!

What is the significance of this result? In computing gauge invariant Green's functions, it is
redundant to consider multiple gauge copies in the path integral.
However, if one is computing the imaginary part of the vacuum energy, the result
depends crucially on the analytic continuation. The analytic continuation can be made gauge 
dependent provided opposite continuations are made only over gauge copies that are not
connected by small gauge transformations. 

This simple resolution of the symmetry problem can be seen in yet another gauge theory,
namely a $U(1)$ gauge theory in $0+1$ dimensions (\ie quantum mechanics). A topological 
gauge boomeron exists much in the same way as sphalerons in $2+1$ dimensions. In this case
the topological requirement is the existence of 
non-trivial maps from the compactified space-time $S^1$ (a circle) to the gauge group
$U(1)$ ($\Pi_1[U(1)] = Z$). Let us explicitly construct the boomeron in a model. The
action is:
\be 
S_E = \int dt \left [ |Dx|^2 + U(|x|) \right ]
\label {uoneaction}
\ee
where $x = {1 \over \sqrt {2}} ({x_1 + ix_2})$ is a complex coordinate and the covariant derivative is
given by $Dx = d_t x -igAx$. The gauge field $A$ has no kinetic term in the action. The potential 
$U$ may be chosen to be of the form shown in Fig.~5a, with a minimum at $|x| = a$. 

\centerline{
\epsfxsize=2.0in \epsfbox{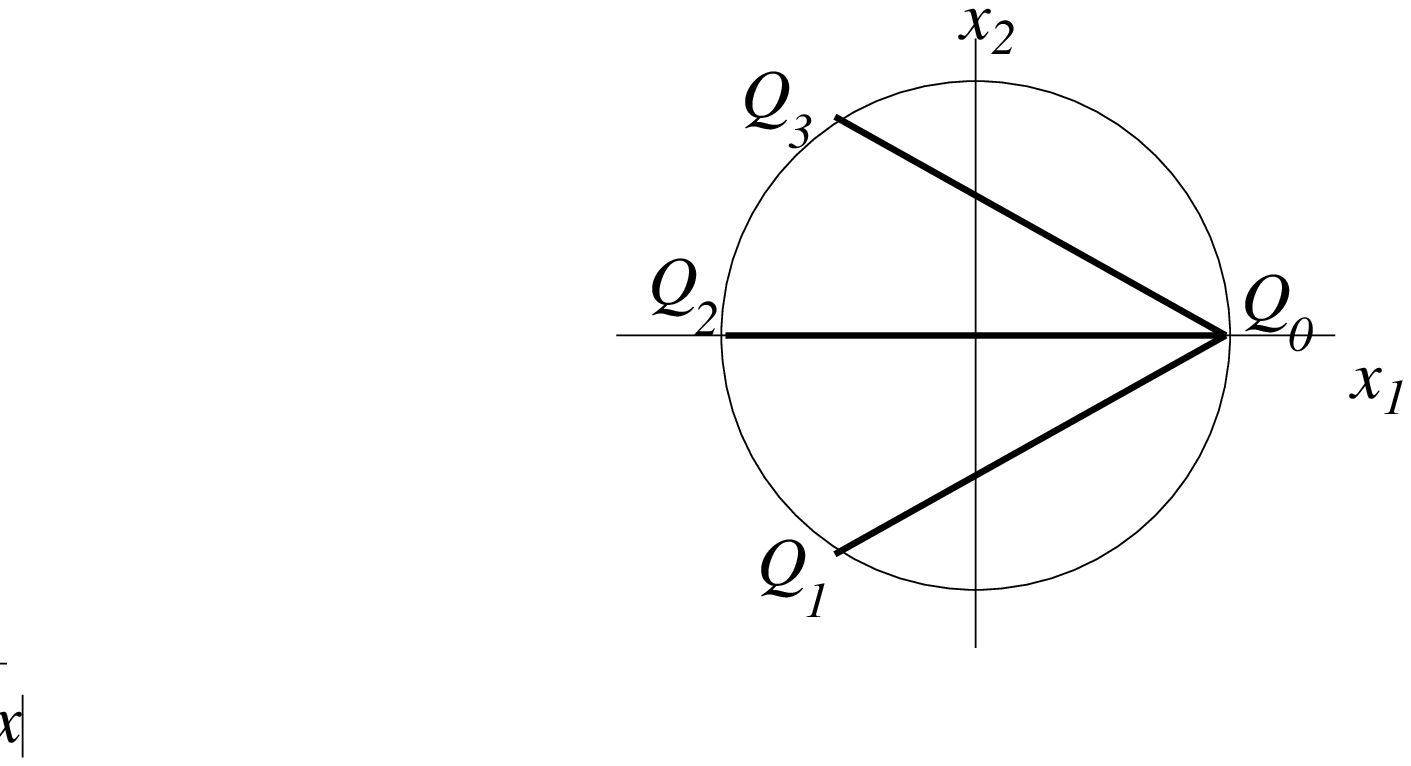}
}

\noindent
\makebox[0.8in][l]{\hspace{2ex} Fig. 5a}
\parbox[t]{4.8in}{ {\small (Left). The potential $U(|x|)$ has a minimum at $|x|=a$.}}\\
\makebox[0.8in][l]{\hspace{2ex} Fig. 5b}
\parbox[t]{4.8in}{ {\small  (Right). The circle 
in the $x_1 - x_2$ plane is defined by $|x|=a$. Bold lines denote trajectories.} } \\

\bigskip

The theory
can be quantized in the axial gauge $A=0$, where Schr{\"{o}}dinger's equations have no
dependence on $A$ and the wave functions are $U(1)$ invariant wave functions of the
corresponding theory with no gauge field. Where are the boomerons? There are only two 
classical solutions of the Euclidean equations of motion that begin from a classical
ground state (say $x=a$) with zero initial velocity. The first is the trivial solution
$x \equiv a$. The second is the solution that starts at $x=a$ and ends at $x=-a$ and
goes along the real line $x_2 =0$. Consider now
trajectories $x(t)$ that begin ($t \to -\infty$) at $x=a$ and end ($t \to +\infty$) 
at $x=a_1+ia_2$ \, with $a_1^2+a_2^2 =a^2$. On the space $P(a, 0; a_1, a_2)$ of all such 
trajectories with fixed $a_1, a_2$,
there is a trajectory $x(a, 0; a_1, a_2)$ that has the least action. Fig.~5b shows three
such trajectories for different values of $a_1$ and $a_2$. As one
moves the end point $x = a_1 + i a_2$ of the trajectory $x(a, 0; a_1, a_2)$ 
clockwise along the circle $|x|=a$ through
the points $Q_0, Q_1, Q_2, Q_3, Q_0$ the value of the action
increases monotonically from its least value $0$ for the trivial trajectory $x(a, 0; a, 0)$
to a maximum at $x(a, 0; -a, 0)$ and then decreases again to $0$ at the trivial trajectory.
Thus the 
solution $x(a, 0; -a, 0)$ to the Euclidean equations of motion is a saddle point of the
action. This trajectory still does not resemble a boomeron because the boundary conditions
at $t \to \pm \infty$ are different. However, that shortcoming can
now be removed by the gauge transformation
$x \to {\rm {exp}}[i\alpha (t)]$, with any $\alpha (t)$ satisfying $\alpha (-\infty) = 0, \alpha (
\infty) = \pi$
and $A(t) = {1 \over g} d_t\, \alpha (t) = A(-t)$. This results in the trajectory 
`closing in' to give
periodic boundary conditions for both $x$ and $A$. 

This is perhaps the simplest gauge-boomeron that posseses properties typical of the
sphaleron. The charge $Q = {g \over 2\pi} \int dt\,  A(t)$ for the boomeron (anti-boomeron) 
is $1/2$ ($-1/2$).
Zero modes of the boomeron from internal symmetries can involve only small gauge
transformations, while the boomeron goes to the anti-boomeron only through a large
gauge transformation. From a boomeron point of view, the cancellation of the imaginary
part of the vacuum energy is identical to the case of the 
sphaleron in the $2+1$ dimensions.

The discussion above brings out the general features of gauge boomerons in arbitrary dimensions. 
However, there may be other physically relevant processes associated with boomerons which
depend crucially on the dimension of space time. In the above $0+1$ dimensional case
for instance, when one chooses the gauge $A=0$, the boomeron has non-periodic
boundary conditions that resemble an instanton. In fact it is easy to see that
these instantons play a physical role: they mix the asymmetric classical ground-states
of the theory to produce a $U(1)$ symmetric ground-state. Such an effect can not be
expected to happen in $2+1$ dimensions, where the gauge fields can not be
completely gauged away and spontaneous symmetry breaking does indeed take place
($SU(2) \to U(1)$). On the other hand, the boomerons in the $2+1$ dimensional theory 
may play a role in Fermion condensation and spontaneous breaking of global chiral symmetries
(suitably defined for $2+1$ dimensions) that has no analog in the $0+1$ dimensional
case.
Suffice it to say here, that there are physical amplitudes that are affected by the boomerons.
Typically, these effects are not related to the 
analytic continuations that yield imaginary integrals. 
They are obtained when one computes the amplitudes by making phenomenological
estimates of the apparently divergent integrals.
We propose to explore some of these aspects in a future publication \cite {dgw} and restrict
ourselves to the study of imaginary contributions to vacuum energy in this paper.

Finally we ask the question:
Do boomerons exist in 4-dimensional gauge theories? A promising hunting ground
would be topological sphalerons of 5-dimensional
theories that arise 
if a nontrivial map exists from the compactified space $S^4$ to
the gauge group space $G$, \ie $\pi_4(G) \ne I$. The following table
from ref.~\cite {table} is useful: 
\be 
\pi_4(G) \simeq \left \{ \begin {array} {rcl} Z_2 \oplus Z_2 \; \; &{\rm
{for}}& \; \; (G = SO(4), Spin(4))  \\ 
Z_2 \; \; \; \; &{\rm {for}}& \; \; (G = Sp(n), SU(2), SO(3), SO(5),
Spin(3), Spin(5)) \\
I \; \; \; \; &{\rm {for}}& \; \; (G = SU(n) (n \ge 3), SO(n) (n \ge 6),
G_2, F_4, E_6, E_7, E_8)
\end {array} \right.
\label {table}
\ee
An $SU(2)$ or $SO(4)$ gauge theory in 4-dimensions therefore should have a 
boomeron- anti-boomeron pair purely on topological grounds. 
The boomeron for the $SU(2)$ gauge theory does in fact seems to exist provided
the action contains some higher- than- dimension- $4$- operators to constrain the 
scale of the solution. This saddle point has been called
the $I^{*}$ instanton in ref.~\cite {istar}.
Physically, the $I^{*}$ may make a noticable
impact on four Fermion scattering amplitudes, to which it contributes a term that
grows exponentially with energy \cite {istar}.
Once again
there do not seem to be any small gauge transformations relating the $I^*$ to its
time- reversal -conjugate.
When $SU(2)$ or $SO(4)$ is
embedded in a larger group, the boomerons
continue to be extrema of $S_E$. The existence of non-topological
boomerons such as these can not be ruled out although
it is not clear if the number of 
negative eigenmodes can still remain 1. 

Since topological boomerons are easier to find
we will remark on another kind arising in 2- dimensional non-linear
sigma models. When
the target space is a manifold $M$ with local coordinates $X^{\mu}$, the
action is:
\be
\label {sigma}
S_E = \int d \tau d \sigma 
\left [  g_{\mu \nu}(X) [(\partial _{\tau} X^{\mu}
\partial _{\tau}  X^{\nu} ) + (\partial _{\sigma} X^{\mu}
\partial _{\sigma}  X^{\nu} ) ] + V(X ) \right ]
\ee
where $\sigma \in [0, 2\pi]$ is the compact space dimension, $\tau$ is time
 and  $g_{\mu \nu}$ is a metric on $M$. The classical ground state 
is the trivial configuration 
($X (\sigma ) \equiv X _0$) where $X_0$ minimizes the potential $V$.
We are interested in field configurations $X(\sigma, \tau)$ which begin and 
end (as $\tau \to \pm \infty$) at the classical ground state
$X(\sigma) \equiv X_0$. The space of these configurations is the space 
$[T^2, M]$ of based maps from a torus $T^2$ to $M$. It can be shown
that the space of loops $[S^1, [T^2, M]]$ on $[T^2, M]$ is isomorphic to the 
space $[T^2, [S^1, M]]$~\cite {algebra}. Therefore $\Pi _1 ([T^2,M])$ is 
non-trivial whenever there are non-trivial homotopy classes of the maps 
$[S^1, M]$ or $[T^2, M]$. 
In particular, topological boomerons should exist in this case if
$M$ is a compact but not simply connected manifold.

\section {Is Time Reversal Invariance Necessary?}

We have shown that the boomeron and the anti-boomeron contributions cancel each
other if the action is time-reversal- (${\cal {T}}$) invariant. One can ask what
happens if $S_E(B) \ne S_E(\overline {B})$? Although the sign of the
contribution is opposite, the anti-boomeron is weighted by a different
exponent than the boomeron and a perfect cancellation may not take
place. Is it possible, then, to devise a situation
where the P.I. formalism fails? 

We will restrict ourselves to the purely bosonic
case. Consider a 4-dimensional field theory first. Let us denote a generic gauge
field by $A$ and a generic scalar field by $\phi$. Both are multiplets
of some representation of the internal symmetry groups. Then there is
only one ${\cal {T}}$ non-invariant renormalizable term that can be
added to the action. It is obtained by contracting the 
field strength tensor with its dual: $S 
= S^T +  \theta \int d^4x [ F \wedge F]$ where $S^T$ is the ${\cal {T}}$
invariant part of the action.
In a 
Q.C.D. like theory the extra term may arise due to instanton effects and
breaks ${\cal {CP}}$ and ${\cal {T}}$. An interesting property of this term is that in the
Euclidean form of the action this term is pure imaginary, \ie, $S_E =
S_E^T + i \theta \int d^4x [ F \wedge F]$.
Under ${\cal {T}}$ this term goes to negative of
itself ($F \wedge F \to - F \wedge F$). Because the Euclidean action is
complex, the equations of motion coming from the real and imaginary parts
of the action must be individually satisfied. The new equation coming
from the imaginary part of the action is 
\be
D\wedge F = 0 \; ,
\label {dfzero}
\ee
which is an identity. Therefore the boomerons and anti-boomerons of the
${\cal {T}}$ invariant theory remain boomerons and anti-boomerons when the new
term is added to the action. The contribution to the vacuum energy density from
a boomeron anti-boomeron pair is (using the field theoretic
generalization of (\ref {multisaddle}))
\be
\delta E_0 = J \times
\left [ ({\rm
{det}}^{\prime}\, O[B])^{-1/2} {\rm {exp}} (i n \theta ) +  ({\rm
{det}}^{\prime}\, O[\overline {B}])^{-1/2} {\rm {exp}} (-i n \theta )
\right ] \; ,
\ee
where $J = 
{\left (-S_E^T(B)/2\pi \hbar\right ) ^2{\rm {exp }}[-S_E^T(B)/\hbar]\over ({\rm
{det}}\,O[\phi])^{-1/2}}$, 
$n = \int d^4x [ F \wedge F]$ is the winding number of the
gauge field in the boomeron background and $\phi$ is the trivial field
configuration corresponding to the vacuum being considered. 
The determinants ${\rm
{det}}^{\prime}\, O[B]$ and ${\rm {det}}^{\prime}\, O[\overline {B}]$ are the same as in the
${\cal {T}}$ invariant theory because the ${\cal {T}}$ non-invariant term is a constant
topological phase factor for the entire Gaussian functional
integration. In particular, the determinant factors are imaginary and 
can be chosen to have opposite signs. Then
the contribution of the boomeron is the complex
conjugate of the contribution of the anti-boomeron and $\delta E_0$ is
real! 

We will give another example. In a
3-dimensional gauge theory the Chern-Simons term $k (A \wedge F + {2 \over
3} A \wedge A \wedge A)$ is a ${\cal {T}}$ non-invariant term. 
In the presence of Higgs fields and if the gauge
group is $SU(2)$, boomerons which are simply the dimensionally reduced
sphalerons, are present in the ${\cal {T}}$ invariant theory ($k=0$). With the inclusion
of the Chern-Simons term ($k \ne 0$)the Euclidean action gets an imaginary
contribution: $S_E = S_E^T + i (C.S)$, where $S_E^T$ is the
${\cal {T}}$ invariant action and $C.S$ is the Euclidean Chern Simons term. The
extra equation of motion to be satisfied by a boomeron is:
\be
F = 0 \; ,
\label {cs}
\ee
Boomerons of the ${\cal {T}}$ invariant
theory do not satisfy (\ref {cs}), and the problem
has an easier resolution than the ${\cal {T}}$ invariant case! There are 
simply no boomerons now!

In general we would like to consider 
non-renormalizable terms as well.
The Euclidean action may be written
as: $S_E = S_E^T + i S_E^U$, where $S_E^U$ is the
${\cal {T}}$ non-invariant part. The boomeron satisfies the equations:
\be
{\delta S_E^T \over \delta \phi} = 0 \; , \nonumber \\
{\delta S_E^U \over \delta \phi} = 0 \; 
\label {general}
\ee
where $\phi $ stands for all the fields in the theory. Under time
reversal these equations are unchanged,
therefore the boomeron $B$ has a corresponding anti-boomeron
$\overline {B}$ which is obtained by time reversing the boomeron. Thus
$S_E[B]$ is the complex conjugate of $S_E[\overline {B}]$. It also
follows that the operator $O[B] = O[\overline {B}]^{\dagger}$,
where $O[B]$ is defined as ${\delta ^2 {\cal {L}}_E \over \delta \phi
^2}[B]$ with ${\cal {L}}_E$ given by
$S_E = \int d^d x [{\cal
{L}}_E]$. 

Now consider the contribution of the boomeron- anti-boomeron pair to the
vacuum energy density in a $d$ dimensional theory:
\bear
\delta E_0 &=&  ({\rm {det}}\, O[\phi])^{1/2}
\left [ (-S_E[B]/2 \pi\hbar)^{d/2} {\rm {exp}}(-S_E[B]/\hbar) 
({\rm {det}}^{\prime}O[B])^{-1/2} \right . \\ \nonumber
 &+& \left .(-S_E[\overline {B}]/2 \pi \hbar)^{d/2}  {\rm {exp}}(-S_E[\overline {B}]) ({\rm {det}}^{\prime}O[\overline
{B}])^{-1/2} \right ] \; ,
\label {finale}
\eear
where $\phi$ denotes the trivial configuration at the vacuum of
interest. In view of our discussion above, the second term on the R.H.S
may be chosen to be the complex
conjugate of the first term and $\delta E_0$ receives no imaginary
contribution from boomerons and anti-boomerons.

The above result is surprising because the ${\cal {T}}$ invariance of the action,
that seemed so vital for the cancellation between boomerons and
anti-boomerons now seems redundant. On the other hand this is a
satisfying result that establishes the independence of the robustness of
the P.I. formalism from the symmetries of the action.

\vspace {2.0cm}
{\centerline {\bf {Acknowledgements}}}

I thank Ryan Rohm for critical readings of the manuscript and Manash Mukherjee for 
many patient primers on algebraic geometry. I am greatly indebted to 
Frans Klinkhamer for some very stimulating discussions.
This work was supported by the Department of Energy under the grants 
DE-FG02-91ER40676 and DE-FG02-84ER40153.

\vfill

\end{document}